\documentclass[letter]{aa} 
\usepackage{graphicx}
\usepackage{float, caption}
\usepackage{rotating}

\usepackage{txfonts}
\usepackage{natbib}   
\usepackage{epstopdf}

\newcommand{\kms}{{km~s$^{-1}$}}  

\newcommand{\vt}{$V_t $}
\newcommand{\vr}{$V_r $}
\newcommand{\vtot}{$V_{tot} $}
\newcommand{\dph}{$d_{ph} $}
\newcommand{\tph}{$t_{ph} $}
\newcommand{\mua}{$\mu_{\alpha} $}
\newcommand{\mud}{$\mu_{\delta} $}
\newcommand{\arcsecyr}{arcsec~yr$^{-1}$}

\begin{document}


\title{HIP 21539 is not a past very close neighbour of the Sun}

\author{F. Crifo
       \inst{1}
\and
C. Soubiran
\inst{2}
\and 
G. Jasniewicz
\inst{3}
\and 
 D. Katz
 \inst{1}
\and 
P. Sartoretti
\inst{1}
\and
P. Panuzzo
\inst{1}
}

\institute{
GEPI, Observatoire de Paris, PSL Research University, CNRS, Univ. Paris
Diderot, Sorbonne Paris Cit\'e, 5 Place Jules Janssen, 92190 MEUDON, France 
\and
Laboratoire d'astrophysique de Bordeaux, Univ. Bordeaux, 
CNRS, B18N, all\'ee Geoffroy Saint-Hilaire, 33615 PESSAC, France
\and
LUPM, UMR CNRS/Universit\'e de Montpellier, CC 72,  Universit\'e de Montpellier, 
34095 MONTPELLIER Cedex 05, France
}


 \abstract
   {}  
{
A previous study  claimed that the star HIP 21539 passed close to the Sun, 
at a distance of 1.9 pc,  around 0.14 Myr ago. We show that this 
is not the case.
}
{ 
We redetermined the trajectory of the star relative to the Sun using a new  
accurate radial velocity from the HARPS spectrograph combined with the recent 
Gaia-TGAS astrometry.
}
{With this new data, the closest approach  of HIP 21539 to the Sun is now 17 
pc, instead of 1.9 pc.
}
{
At this distance, the star has not perturbed the Oort cloud.
}


\maketitle


\section{Introduction}

We present a short improvement to the interesting paper by \cite{dyb15}
about close stellar passages at less than 2 pc from the Sun with a
possible impact on cometary orbits.
As part of the work aimed at defining stellar radial
velocity standards (RV-STD) for the calibration of the Gaia Radial
Velocity Spectrometer (RVS)  
\citep[]{Crifo10, soub13}, 
we obtained a much better radial 
velocity (RV) for one of their targets, i.e. \textbf{HIP 21539.} This new RV is 
derived from three independent observations by the HARPS spectrograph 
and rules out the possibility 
for this star to have had such a close passage in the recent past.
New values for the date and minimum distance are estimated with the 
straight line approximation.

\section{The problem}

\cite{dyb15} carefully examined the possible candidates for
close passages of already nearby stars at a distance less than 2 pc from
the Sun, as such objects may strongly perturb the Oort cloud.
Just before, \cite{bj15} carried out a very similar work. 
Both papers used the XHIP catalogue by \cite{xhip} as an entrance list.
The XHIP catalogue contains all necessary data. Parallaxes and proper
motions are taken from the HIP-2 catalogue \citep[]{hip2}, but radial velocities 
come from a vast compilation made by Anderson and Francis, who 
really searched deep in the literature for all possible existing data.

The star HIP 21539 is found only in the paper by \cite{dyb15}; it is supposed
to have had its closest approach 0.14 Myr ago at a perihelion distance of 
1.92 pc. The corresponding radial velocity is 248 km s$^{-1}$, issued from 
the \cite{barbier94} catalogue, itself referring to \cite{contreras70}, 
\textit{``Radial velocities for twenty-three stars selected from an objective prism 
survey are communicated. The data indicate that the peculiar G- and K-stars 
included in the program constitute a high velocity group''}. The data quality 
is quoted as ``D'' in XHIP, i.e. the lowest quality. No other value is available
in Simbad or Vizier.

\section{New radial velocity}

In order to find additional RV-STD for the Gaia RVS, we searched 
the AMBRE-HARPS catalogue \citep{ambre14}. 
This catalogue provides atmospheric parameters for the ESO:HARPS 
archived spectra, together with radial velocities either derived by 
the ESO:HARPS reduction pipeline or by the AMBRE pipeline.
The HARPS spectrograph is a velocimeter mounted on the ESO 3.6m telescope at La Silla, with a 
resolving power of $ R~=~\lambda /\Delta \lambda$ = 115000; for more
details see \cite{HARPS}.

To be consistent with our previous lists of RV-STD candidates  
\citep[]{Crifo10,soub13}, the RV 
measurements must be expressed in the SOPHIE scale; SOPHIE is another 
velocimeter at Observatoire de Haute-Provence with R~=~75000 and a
reduction pipeline similar to that of HARPS.

For HIP 21539, the AMBRE-HARPS catalogue provides three measurements of RV 
from the ESO pipeline at dates 2003/12/11, 2004/02/01, and 2004/11/26, 
i.e. a time span of 351 days.

The weighted average of these three values gives
\begin{center}
 \textbf{RV = 26.926 \kms;  $\sigma_{\rm RV}$ = 0.0026 \kms .}
\end{center}

The star was integrated into the list of radial velocity standards for the RVS.

\section{Linear approximation}

We now calculate a new approximate minimum distance and corresponding date for 
the closest approach of the star to the Sun, called \dph~ and \tph~ (distance 
 and time from perihelion) according to \cite{bj15}.
\citeauthor{bj15} and \citeauthor{dyb15}  first computed  \dph~ and 
\tph~  with 
the linear approximation and then with the introduction of the Galactic 
potential  perturbing the linear motion; \citeauthor{bj15} used the linear
approximation for a first gross 
selection within the XHIP catalogue \citep[]{xhip}. In their Table 2, 
\cite{dyb15} compare the results of the two calculations: the difference 
for \dph~ becomes noticeable for |\tph|~>~3Myr, i.e. for far-away stars that 
travelled  long enough in time to have felt the influence of 
the Galactic potential.
In our case, the linear approximation is largely sufficient; moreover the 
final value  of \tph~ ($\sim$~1~Myr) shows that it is appropriate.

\begin{figure}[ht!]
 \centering
 \includegraphics[width=9cm]{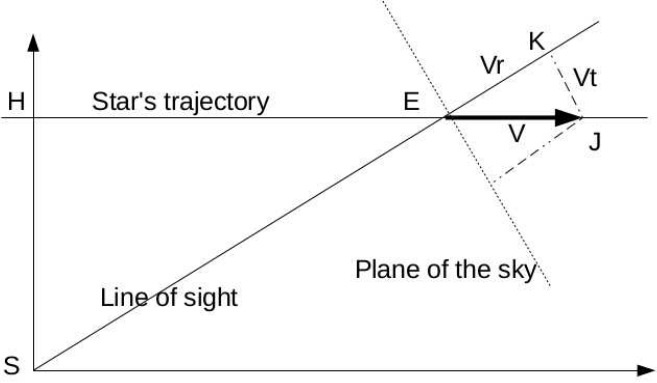} 

  \caption{\textit{Trajectory of the star relative to the Sun. The star E 
  moves uniformly along the straight line. H is the perihelion.}}
  \label{crifo:fig1}
\end{figure}

Figure 1, taken from \cite{green}, fig. 11.1, illustrates the positions:
S = Sun; E is the star at our epoch (SE = $r$), and assumed to move with an 
uniform velocity along a straight line; H is the closest approach to the Sun
(perihelion for Bailer-Jones), at a distance SH = \dph~ and a time \tph~
to be calculated (origin of time at E).
The total velocity $\vec{V}$ is projected over the line-of-sight SE and 
the plane of the sky; the components are \vr~ and \vt.

In the right triangle EKJ we have EK = \vr; JK = \vt.
By comparing the right triangles EKJ and EHS, we may write

SH/SE = JK/JE = \vt~ / \vtot~ = \dph~/$r$

HE/SE = \tph~ .\vtot~/$r$ = KE/JE = \vr~ /\vtot
.

Hence, \\

\dph~ = $r$. \vt~/\vtot; 
\tph~ = -$r$.\vr~ /{${V_{tot}}^2.$} \\

A sign ``-'' must be introduced in front of the expression of \tph~ , as 
the origin of time is supposed to be at position E, and \vr~ is positive 
when the star is receding from the Sun.

The values \vt~ and \vtot~ are calculated from the proper motion 
components $\mu_{\alpha}$
and $\mu_{\delta}$ and the parallax $\varpi$,\\

$V_t = k.{\sqrt{{\mu_{\alpha}^2} +{\mu_{\delta}}^2}} /{\varpi}$ ~;
$V_{tot} = \sqrt{{V_t}^2 + {V_r}^2}$,\\

where $k$~= 4.74  is the coefficient converting the \arcsecyr~ in \kms 
(see \citeauthor{bj15} eq. 5; \citeauthor{green}, eq. 11.8).

\section{Results}

The resulting numerical values are given in Table 1 in the following three 
cases:

- Cases ``old'' and ``new'' for the two values of RV: the old 
bad
value and the 
new HARPS value, combined with the parallax and proper motion 
from Hipparcos-2, as in \citeauthor{dyb15}, table 2.

- Case ``TGAS'' for the HARPS RV combined with the recently published parallax 
and proper motion from the TGAS Catalogue:
Tycho-Gaia subset, available at CDS \citep{tgas}.

In Table 1, each ``case'' is made of two lines: the upper line contains 
the data itself; the lower line (noted ``sig'') contains the corresponding 
errors, taken from SIMBAD (HIP2) or TGAS. An arbitrary error of 20 \kms~ was 
adopted for the old RV of 248 \kms (order of magnitude for error on radial 
velocities obtained with objective prism, but unrealistic here).
Using the TGAS data instead of HIP-2 improves the accuracy of
\dph~ and \tph.

With updated RV, parallax and proper motion, the closest approach of 
HIP 21539 to the Sun is now 17.3 pc, $\sim$1 Myr ago instead of 1.9 pc 
and 0.14 Myr ago, as first computed by \citeauthor{dyb15}.

\begin{table}[!h]

\caption{Calculation of old and new distance of perihelion}
\centering 
\begin{tabular}{l r r r r r r}
 \hline\hline
 case   & $\varpi$ &  \mua   &    \mud  &    \vr   &   \dph &   \tph \\
        &  mas  &  mas.yr$^{-1}$ &  mas.yr$^{-1}$  &  \kms  &  pc  &   Myr \\
  
  \hline 
  
old    &  28.580 &  -80.73 &   15.70 &   248.000 &   1.92 &  -0.14 \\ 
sig    &   1.340 &    1.08 &    1.28 &    20.000 &   0.20 &   0.04 \\
 
\\

new    &  28.580 &  -80.73 &   15.70 &    26.926 &  15.81 &  -1.03 \\ 
sig    &   1.340 &    1.08 &    1.28 &     0.003 &   1.06 &   0.05 \\ 

 \\
 
TGAS &  27.270 &  -81.31 &   15.27 &    26.926 &  17.27 &  -1.06 \\ 
sig  &   0.280 &    0.08 &    0.09 &     0.003 &   0.25 &   0.01 \\ 
\hline
\end{tabular}

\end{table}

\section{Conclusions}

 These new data show that HIP 21539 did not pass very close to
the Sun; it therefore certainly did not perturb the Oort cloud.
This short paper shows the importance of reliable RV for a good
description of the solar neighbourhood and Galactic mechanics. The RVS 
on board Gaia is expected to provide radial velocities for 
more than 100 millions stars and revolutionize our knowledge of 
kinematics in the solar neighbourhood.
%

%
%

\begin{acknowledgements}

Many thanks to the CDS, particularly for the SIMBAD database and the
Vizier section with so many old catalogues and data that are carefully 
stored and made available.

This work has made use (for Table 1) of data from the European Space 
Agency (ESA)
mission {\it Gaia} (\url{http://www.cosmos.esa.int/gaia}), processed by
the {\it Gaia} Data Processing and Analysis Consortium (DPAC,
\url{http://www.cosmos.esa.int/web/gaia/dpac/consortium}). Funding
for the DPAC has been provided by national institutions, in particular
the institutions participating in the {\it Gaia} Multilateral Agreement.

\end{acknowledgements}

\bibliographystyle{aa}  
\bibliography{biblio} 

\end{document}